\documentclass[aps, prl, reprint ]{revtex4-1}

\pdfoutput=1 
\usepackage{graphicx}
\usepackage{amssymb, amsmath, amsthm, amsfonts, bm, mathrsfs, wasysym, txfonts, bbm, enumerate, color}

\begin{document}

\title{{\large Axionic excitation from QCD Condensate}}

\author{Chi Xiong}
\email[]{xiongchi@ntu.edu.sg}
\affiliation{${}^1$Institute of Advanced Studies, \\ ${}^2$School of Physical and Mathematical Sciences, \\Nanyang Technological University, Singapore 639673 }


\begin{abstract}
A complex order parameter is used for describing inhomogeneous quark condensate and its phase is decomposed into a regular part and a singular part. An emergent gauge symmetry is found to connect these two parts. The singular part plays the role of an emergent gauge field while the regular part can be considered as an axionic excitation, provided that the chromoelectric flux tube is capable of inducing a vortex configuration in the quark condensate, with the help of axial anomaly.

\vspace{0.2cm}
\end{abstract}

\maketitle               

\section*{Introduction}

The vacuum structure of Quantum chromodynamics (QCD) is extremely complicated. However, one of characterizing features of the QCD vacuum, quark condensate, is usually treated as  a constant. In quantum fluids such as Bose-Einstein condensate (BEC) of cold atoms, superfluid helium and the Cooper pairs of electrons in superconductors,  condensates are described by macroscopic wave functions or order parameters, which satisfy certain nonlinear differential equations such as the Gross-Pitaevskii equation and the Ginzburg-Landau equation. Therefore it is natural to consider the possibility of inhomogeneous quark condensate under certain circumstances. For instance, the chiral condensate may be spatially modulated at high densities (see Ref. \cite{Buballa:2014tba} for a review). In some flux-tube models of hadrons \cite{Suganuma}, the chiral condensate is considered strongly distorted near the hadron and vanishes inside the hadron, which suggests a restoration of chiral symmetry in there, provided that the chromoelectric field is strong enough to pull the quark-antiquark pairs apart.

In this letter we treat quark condensate as a complex order parameter and allow it to be space and time dependent. Elementary excitations might be produced in the condensate. We consider one particular type of such excitations, i.e. quantum vortices and associate them with the QCD flux-tube (or string) model of hadrons. We show that such a configuration can produce axionic excitation which  is coupled to the Chern-Simons current. A phenomenological Lagrangian is proposed for such a flux-tube model of pseudoscalar mesons. 

\section*{Emergent gauge field for topological defect}

We introduce a complex scalar field operator 
\begin{equation}
\hat{\Phi}(x) = \Phi(x) + \hat{\varphi}(x)
\end{equation}
to describe the quark condensate and the particle excitations. The macroscopic wave function or order parameter 
\begin{equation} 
\Phi (x)  \equiv F (x) \,e^{i \sigma (x)} 
\end{equation}
is a complex function and $F(x)$ and $\sigma (x)$ are the magnitude and the phase of $\Phi$, respectively. They are related to the quark condensate as
\begin{equation}
\left\langle \bar{q}^i_R q^j_L \right\rangle = - \Phi (x) \, \delta^{ij} = - F (x) \,e^{i \sigma (x)} \, \delta^{ij}
\end{equation} 
Note that here we do not make the assumption that the QCD vacuum has a definite parity, hence in general $\Phi$ is complex, similar to the cases of BEC, superfluidity and superconductivity in which the order parameters are complex functions. The chiral symmetry group is ${SU}_V(N_f) \otimes {SU}_A(N_f) \otimes U_V(1) \otimes U_A(1) $ and $N_f$ is the number of light quark flavors. 
Under the $U_A(1)$ transformation the quark fields transform as 
\begin{equation}
q_L \rightarrow e^{ - i \xi /2} q_L, ~~~q_R \rightarrow e^{ i \xi /2} q_R, 
\end{equation}
hence the scalar fields $\Phi$ and $\Phi^*$ transform as
\begin{equation}
\Phi \rightarrow e^{- i \xi} \Phi, ~~~ \Phi^* \rightarrow e^{ i \xi} \Phi^*
\end{equation}
so one can see that $\Phi$ does transform under the $U_A(1)$ symmetry while stays invariant under the $U_V(1)$ symmetry. We will show later the importance of introducing a non-vanishing phase field $\sigma(x)$. For simplicity we also take one value for different quark condensates while in general we should treat the $\Phi$ field as a matrix (like the chiral field in the chiral perturbation theory). Suppose that as an order parameter the function $\Phi$ is space and time dependent (due to external sources) and described by the Lagrangian density
\begin{equation} \label{L0}
\mathcal{L}_{0}(\Phi, \Phi*)= - (\partial^{\mu}+ i S^\mu) \Phi^{\ast} (\partial_{\mu} - i  S_\mu) \Phi - V\left(\Phi^{\ast}, \Phi\right) 
\end{equation}
where $S_\mu$ is some external gauge field or source (e.g. those external currents used in the chiral Lagrangian approach), and $V\left(\Phi^{\ast}, \Phi\right) $ a nonlinear potential of $\Phi \Phi^*$. We begin with a decomposition of the derivative of the phase of $\Phi$,
\begin{equation} \label{decomp}
\partial_{\mu}\sigma=\partial_{\mu}\vartheta + X_{\mu}
\end{equation}
where the scalar $\vartheta (x)$ is the smooth part of the phase function $\sigma$ and the vector $X_\mu$ the singular part.  For topological defects like vortices, a more rigorous definition for the ``smoothness" and ``singularity" is 
\begin{equation}
[\partial_\mu,  \partial_\nu] \,\vartheta = 0, ~~\textrm{while} ~[\partial_\mu,  \partial_\nu] \,\sigma = \partial_\mu X_\nu -  \partial_\nu X_\mu \equiv X_{\mu\nu} \neq 0.
\end{equation}
hence, the tensor $X_{\mu\nu}$ measures the vorticity associated with the singularity. Note that there is a ``symmetry" in the decomposition (\ref{decomp}) -- the phase $\sigma$ is invariant under the transformation 
\begin{equation} \label{eg}
\vartheta \rightarrow \vartheta + \alpha, ~~X_\mu \rightarrow X_\mu - \partial_\mu \alpha
\end{equation}
where $\alpha$ is arbitrary smooth function. Interestingly, this simple ``symmetry" leads to interpretation of the vector field $X_\mu$ as an emergent gauge field, based on the following observation: If we define a new order parameter, $\phi$, with $\vartheta$ being its phase and $F$ its amplitude,
\begin{equation}
\phi \equiv F (x) \,e^{i \vartheta (x)}
\end{equation}
then the original Lagrangian (\ref{L0}) can be rewritten in terms of $\phi$ and $X_\mu$,
\begin{equation} \label{L0g}
\mathcal{L}_{0} =\left(  \partial^{\mu}+ i S^{\mu} - i X^\mu\right)  \phi^{\ast
} \, (  \partial_{\mu}-i S_\mu + i X_{\mu} )  \phi+V\left(  \phi^{\ast}\phi\right)
\end{equation}
with an emergent gauge symmetry
\begin{align} \label{emergent}
\phi \rightarrow \phi^{\prime}  &  =e^{i\alpha}\phi \nonumber\\
X_{\mu}\rightarrow X_{\mu}^{\prime}  &  =X_{\mu} - \partial_{\mu}\alpha.
\end{align}
One can impose extra constraint(s) on the decomposition (\ref{decomp}) to add other feature(s) such that it will resemble Helmholtz decomposition or things like that. In that case the extra constrain(s) will play the role of ``gauge fixing" for the transformation (\ref{emergent}). 

All the steps above can proceed without the presence of the external source field $S_\mu$. The emergence of the ``gauge" field $X_\mu$ seems to suggest that a theory with global symmetry of field $\Phi$ in (\ref{L0}) is equivalent to another theory with local gauge symmetry of fields $\phi$ and $X_\mu$ in (\ref{L0g}) (taking $S_\mu = 0$ in both cases). Nevertheless, the seemingly trivial procedure above (as it is similar to applying the gauge principle to extend global symmetries to local ones), functions for the purpose of separating singularities of the theory (e.g. the cores of the vortices in present case) from the regular part of the phase. Such singularities occur in real physical systems, e.g. superfluid ${}^4$He and BEC of cold atoms, where quantum vortices appear in these quantum fluids so one has to distinguish the difference between rotational flow and potential flow, and the difference between turbulent flow and laminar flow in more complicated circumstances.  For quantum vortices, $X_\mu$ is further restricted by the quantized circulation condition,  
\begin{equation}
\oint\limits_{C} dx^{\mu}X_{\mu}=2\pi n\text{ \ \ }\left(  n=0,\pm1, \pm2,\ldots\right)
\end{equation}
As mentioned earlier the $U_A(1)$ symmetry of the model (\ref{L0}) leads to an axial current 
\begin{equation}
J^A_{\mu} \equiv  -\frac{i}{2}\left( \Phi^{\ast}\partial_{\mu}\Phi-\Phi\partial_{\mu}\Phi^{\ast}\right) = F^2 \partial_{\mu} \sigma
\end{equation}
At classical level, it is easy to see that from (\ref{L0}) $J^A_\mu$ is conserved.  In terms of variables $\theta$ and $X_\mu$,  the conservation of $J^A_\mu$ leads to (assuming an approximately constant modulus $F \approx F_0$ = const.)
\begin{equation}
\partial^\mu X_\mu = - \Box \, \vartheta \equiv - \partial^\mu \partial_\mu \vartheta
\end{equation}
which vanishes if we choose a gauge condition $\Box \vartheta = 0$ for the transformation (\ref{eg}). Later we will see both conservation laws are broken by the axial anomaly.

Defining $\mu^2 = - \partial^\tau \sigma \partial_\tau \sigma$, we introduce a 4-velocity 
\begin{equation}
U_\mu = \frac{1}{\mu} \partial_\mu \sigma, ~~~~~ U^\mu U_\mu = -1
\end{equation}
and $\mu$ plays the role of chemical potential \cite{XC14}. 
The stress-energy tensor can be put in the usual hydrodynamical form
\begin{equation}
T_{\mu\nu} = T_{\mu\nu}^F + \epsilon U_\mu U_\nu + p (g_{\mu\nu} + U_\mu U_\nu)
\end{equation}
where $\epsilon \equiv \mu^2 F^2 + V(F^2)$ and $p \equiv \mu^2 F^2 - V(F^2)$ are the (total) energy density and the pressure of the ``fluid", respectively, and the tensor $T_{\mu\nu}^F \equiv 2 \partial_\mu F \partial_\nu F - g_{\mu\nu} \partial^\tau F \partial_\tau F$  vanishes for constant $F$. One can compute other hydrodynamics equations (see e.g. cite) and quantities like twist tensor, shear tensor and expansion scalar. Note that with the vorticity tensor  $X_{\mu\nu} = \partial_\mu X_\nu -  \partial_\nu X_\mu$ one can build the vorticity scalar 
\begin{equation}
\omega^2 \equiv X^{\mu\nu}  X_{\mu\nu} 
\end{equation}
which may work as the kinetic term for $X_\mu$ if we consider it as an emergent gauge field and $X_{\mu\nu}$ its field strength.

\section*{Axionic QCD string and anomaly-inflow mechanism}

To describe an axionic QCD string we begin with the following effective Lagrangian,
\begin{eqnarray} \label{gNJL}
\mathcal{L}^0_{\textrm{eff}} &=& -\frac{1}{4} \kappa \, G^a_{ \mu\nu} G^{a\mu\nu} - \partial^{\mu} \Phi \partial_{\mu} \Phi^* - V(\Phi, \Phi^*)  \cr
&+& \bar{q} \left[ i \gamma^{\mu}(\partial_{\mu}  - i g G_{\mu}^a T^a) - g_Y ( \Phi_1 +  i \gamma^5 \Phi_2 ) \right] q \cr
&&
\end{eqnarray} 
where $\Phi_1, \Phi_2$ are the real and imaginary parts of $\Phi$, respectively and the function $\kappa = \kappa (|\Phi|)$ describes the color electric and magnetic polarization properties of the vacuum similar to the (soliton) bag or flux-tube models. For example, $\kappa (F) = 1- F/F_0$ for the soliton bag model.  For a flux-tube or string configuration, 
we first need to check the boundary condition of $\Phi$. In order to describe, for instance a single straight vortex-line configuration with winding number $m \in \mathbb{Z}$, one takes the amplitude and the phase of  $\Phi$ to be, respectively,
\begin{equation} \label{alpha-theta}
F = F(\rho), ~~~~\sigma = m \theta, ~~ m = \pm 1, \pm 2, \cdots 
\end{equation}
with boundary condition 
\begin{equation} \label{vortexbc}
F (0) = 0, ~~~~\textrm{and} ~~ F(\infty) \longrightarrow F_0 \,(=\textrm{constant}),
\end{equation}
where $\rho$ and $\theta$ are the polar coordinates describing the cross section of the straight vortex line.  The condition on $F(\infty)$ can come readily from the quark condensate value, and the question is whether $F(0) = 0$ can be satisfied inside the flux tube.  This has been studied before in the context of the chiral symmetry restoration \cite{Suganuma}. 
The result is that when the chromoelectric field exceeds some critical strength,  (see e.g. a numerical study by Suganuma and Tatsumi \cite{Suganuma}:  $E_{crit} \sim 4 $GeV/fm, $E \approx 5.3 $ GeV/fm $>E_{crit}$ ), the chiral symmetry is restored inside the QCD string. The chromoelectric field in the flux tube, if strong enough, can pull quark and anti-quark pairs apart and hence, destroy the quark condensate.  
Therefore it is possible that the quark condensate vanishes inside the flux tube, $\langle \bar{q}q \rangle = 0 $, while outside the flux tube the QCD vacuum has non-vanishing $\langle \bar{q}q \rangle \neq 0 $. This shows the quark condensate may have the boundary condition of a  vortex configuration. We then assume that the vortex configuration is energetically favored,  as suggested by the appearance of the Abrikosov vortices in the type-II superconductors in an external magnetic field.

With this vortex configuration of the quark condensate $\Phi$ and the flux-tube model for hadrons \cite{XC}, the effective Lagrangian (\ref{gNJL}) now is capable of describing an axionic QCD string. One can study the Dirac equation for the quarks with both $\Phi$ and the gluon potential $G_{\mu}$ being treated as background fields, then what follows is similar to the anomaly-inflow scenario proposed by Callan and Harvey \cite{Callan-Harvey}. It is easy to see that chiral zero modes of quarks are localized in the vortex, since they have an exponential profile $\psi_L = \chi_{L} \, \exp \big[- \int_0^{\rho} F(\rho') d\rho' \big]$ where $\chi_{L}$ is a two-dimensional spinor. The chiral zero modes are also coupled to the gluon potential $G_{\mu}$, therefore a gauge anomaly appears in the vortex, which is cancelled by an effective action \cite{Callan-Harvey}
\begin{equation} \label{cseff}
S_{\textrm{\tiny{C-S}}} = -\frac{ g^2}{16 \pi^2}\int d^4 x \, \partial_{\mu} \sigma \, K^{\mu}
\end{equation}
This cancellation happens  because the massive quark modes which live off the vortex mediate an effective interaction between the quark condensate and the gluon field, which induces a vacuum current 
\begin{equation}
\langle J_{\textrm{\tiny{ind}}}^{\mu a} \rangle =  \frac{g^2 N_f}{8 \pi^2} \epsilon^{\mu\nu\rho\tau} \partial_{\nu} \sigma G^{a}_{\rho\tau}
\end{equation}
Converting it to an effective action one obtains (\ref{cseff}). With this topological term included, the effective Lagrangian becomes
\begin{eqnarray} \label{eff1}
\mathcal{L}^1_{\textrm{eff}} =& - & \kappa (F) /4  ~ G^a_{ \mu\nu} G^{a\mu\nu} - \partial^\mu F \partial_\mu F  - V(F^2)  \cr
&+& \bar{q} \left[ i \gamma^{\mu}(\partial_{\mu}  - i g G_{\mu}^a T^a) - g_Y F e^{ i \sigma \gamma^5}  \right] q - F^2 \partial^\mu \sigma \partial_\mu \sigma  \cr
&- & (g^2/16\pi^2) \, \partial_{\mu} \sigma \, K^{\mu}
\end{eqnarray} 
Due to the decomposition (\ref{decomp}), the last term of (\ref{eff1}) can be rewritten as,
\begin{equation} \label{XK}
 \partial_{\mu} \sigma \, K^{\mu} = \partial_{\mu} \vartheta \, K^{\mu} + X_\mu \, K^{\mu}
\end{equation}
It is tempting to consider the $\vartheta$ field as an axion since it is a well-defined smooth function. After integration by parts, the first term becomes the familiar topological charge term,
\begin{equation}
\partial_{\mu} \vartheta \, K^{\mu} \sim \vartheta \, \partial_\mu K^\mu \sim \vartheta \, G^a_{ \mu\nu} \tilde{G}^{a\mu\nu}
\end{equation}
while the second term $X_\mu \, K^{\mu}$ plays a crucial role in cancelling the lower dimensional gauge anomaly on the axionic string, which has been demonstrated in \cite{Callan-Harvey}. Interestingly, $X_\mu$ contains no derivative and the term  $X_\mu \, K^{\mu}$ cannot be integrated by parts in the same way as the $\partial_{\mu} \vartheta \, K^{\mu} $ term does. However, when studying its gauge variation, one can integrate by parts in the ``opposite" way such that another derivative acts on $X_\mu$ (or $\partial_{\mu} \sigma$) in the form of
\begin{equation}
[\partial_\mu, ~\partial_\nu] \sigma = X_{\mu\nu} \neq 0,
\end{equation}
The mathematical structure of this cancellation is the descent equation \cite{DE}, which relates the chiral anomaly in (2n+2)-dimensions to the gauge anomaly in 2n-dimensions. Writing the the terms in the coupling (\ref{XK}) in differential forms
\begin{equation}
\partial_\mu \sigma dx^\mu \rightarrow d\sigma,  \partial_\mu \vartheta dx^\mu \rightarrow d\vartheta, \\ X_\mu dx^\mu \rightarrow X, ~K^\mu \rightarrow \mathcal{K}^0_{3}
\end{equation}
where $ \mathcal{K}^0_{3}$ is the Chern-Simons 3-form. Under a gauge transformation of $SU(3)_C$ group
\begin{eqnarray}
\delta\int_{M} d\sigma \wedge \mathcal{K}^0_{3} &= & \int_{M} d\sigma \wedge d \mathcal{K}^1_{2} \cr
&=& - \int_{M} d^2\sigma \wedge  \mathcal{K}^1_{2} \cr
&=& - \int_{M} dX \wedge  \mathcal{K}^1_{2} 
\end{eqnarray}
where we have used the identity $d\sigma = d\vartheta + X$ and the smooth condition of the function $\vartheta (x)$, i.e. $d^2 \vartheta = 0$.
The Chern-Simons 3-form $ \mathcal{K}^0_{3}$ is connected to the lower dimensional gauge anomaly $\mathcal{K}^1_{2} $ through
\begin{equation}
\delta \mathcal{K}^0_{3} = d\, \mathcal{K}^1_{2}
\end{equation}
 which is one of the descent equations $\delta \mathcal{K}^{i-1}_{2n-i} = d\, \mathcal{K}^{i}_{2n-i-1} $ \cite{DE} and for a straight vortex with winding number $m=1$ sitting on the $z$-axis (denoted by $\Sigma$), its vorticity is given by
 \begin{equation}
 dX = 2 \pi \delta(x) \delta(y) dx \wedge dy
 \end{equation}
Thus we obtain the counter term for the lower dimensional gauge anomaly
\begin{equation}
\delta\int_{M} d\sigma \wedge \mathcal{K}^0_{3} = - \int_{\Sigma} \mathcal{K}^1_{2},
\end{equation}
such that the whole theory contain no gauge anomaly. 

Noticing that the quarks zero modes are also coupled to the electromagnetic field $A_\mu$ and they lead to a gauge anomaly with respect to the electromagnetism as well. The same anomaly inflow mechanism applies, hence we have another effective coupling term  
\begin{equation} \label{mcs}
\mathcal{S}_{\textrm{\tiny{MCS}}} = -\frac{e^2}{8 \pi^2} \int d^4x \, \partial_{\mu} \sigma \, K^{\mu}_{\textrm{\tiny{MCS}}}
\end{equation}
where $ K^{\mu}_{\textrm{\tiny{MCS}}} =  \epsilon^{\mu\nu\rho\tau} A_{\nu} F_{\rho\tau} $ is the Maxwell-Chern-Simons current. Including all relevant terms we obtain
\begin{eqnarray} \label{eff}
\mathcal{L}_{\textrm{eff}} =& - & \kappa (F) /4  ~ G^a_{ \mu\nu} G^{a\mu\nu} - 1 /4  ~ F_{ \mu\nu} F^{\mu\nu}- \partial^\mu F \partial_\mu F  - V(F^2)  \cr
&+& \bar{q} \left[ i \gamma^{\mu}(\partial_{\mu}  - i g G_{\mu}^a T^a - i e A_\mu)  -g_Y F e^{ i \sigma \gamma^5}  \right] q  - F^2 \partial^\mu \sigma \partial_\mu \sigma \cr
&- & (g^2/16\pi^2) \, \partial_{\mu} \sigma \, K^{\mu} - (e^2/8 \pi^2) \, \partial_{\mu} \sigma \, K^{\mu}_{\textrm{\tiny{MCS}}}
\end{eqnarray} 
which completes our phenomenological model of an axionic QCD string. 

\section*{Applications and discussions}

For ultra-relativistic collision of heavy nuclei, net color charges develop and are connected by  flux tubes with collinear color electric and magnetic field. For example, in \cite{Lappi06} longitudinal color electric and magnetic fields are produced after high energy hadronic collisions. Such gauge configurations carry topological charges ($\vec{E} \cdot \vec{B} \neq 0)$ and lead to net chiral quark zero modes in the flux tubes. An interesting phenomenon is  the chiral magnetic effect (CME) \cite{Kharzeev07, Fukushima08} - charge separation can occur in  a background (electromagnetic) magnetic field, and consequently, an electromagnetic current is generated along the magnetic field.  Our effective Lagrangian (\ref{eff}) can be used to study these phenomena as well, e.g. one can readily derive a modified Maxwell equation based on (\ref{eff})
\begin{equation}
\partial_{\mu} F^{\mu\nu} = J^{\nu} - \frac{e^2}{2 \pi^2} \, \partial_{\mu} \sigma \, \tilde{F}^{\mu\nu}
\end{equation}
and the induced current $\vec{J} \sim -\dot{\sigma} \vec{B}^{\textrm{\tiny{M}}}$ agrees with \cite{Kharzeev07, Fukushima08}.

The vortex configuration and the effective coupling $\partial_{\mu} \sigma \, K^{\mu}$ also allows us to address the $U_A(1)$ problem \cite{Weinberg:1975} in a similar way as it has been done in Ref. \cite{TX1, TX2} with membrane (domain wall) configuration. The essence is the localization of chiral zero modes on topological defects such as vortices and domain walls. The instanton approach \cite{'tHooft:1976} and its extensions (e.g. instanton gas and liquid \cite{Schafer:1996}) also have this feature. Once the Chern-Simons current $K^\mu$ is produced, one can use the anomalous Ward identities \cite{Christo} for singlet and the mixing of the ``ghost pole" of the correlator $\langle K^\mu K^\nu \rangle$ \cite{Luscher78} and the Goldstone pole lead to the correct $\eta'$ mass.  One may speculate that the vortex excitation considered in this letter, reflects the collective motion in the quark condensate and cost certain amount of energy which contributes to the $\eta'$ mass, while for other octet pseudoscalar mesons such vortex excitations do not occur, otherwise their associated axial currents would have been affected by the anomaly as we have shown from the above axionic string model.  
Moreover, flux-tube models seem to have the advantage of directly addressing the confinement problem, so we formulate our model based on the flux-tubes.
The decomposition $\partial_{\mu}\sigma=\partial_{\mu}\vartheta + X_{\mu}$ can be generalized to treat other singularities or topological defects. 

We close by pointing out that if there are several quark condensates separated by some junctions, each of them has a different phase, resembling the Josephson effect in superconductors with Josephson tunnelling junctions. The Josephson critical current $\sim \cos(\Delta \theta)$ may correspond to some axion-potential $ \sim  \cos(\Delta \sigma)$ in QCD  where $\Delta \theta$ and $\Delta \sigma$ are their phase differences, respectively.  It should be mentioned that similar idea has appeared long time ago in Ref.\cite{Minkowski}.

\end{document}